\documentclass[twocolumn]{aastex62}
\newcommand{\Mpch}{Mpc $h^{-1}$}
\newcommand{\Msun}{$\mathrm{M}_{\odot}$ $h^{-1}$}
\newcommand{\ergs}{erg s$^{-1}$}

\usepackage{amsmath,amstext}
\shorttitle{AGN clustering}
\shortauthors{Powell et al.}
\begin{document}

\title{The Clustering of X-ray Luminous Quasars}

\author{M. C. Powell}
\affiliation{Yale Center for Astronomy and Astrophysics, and Physics Department, Yale University, PO Box 2018120, New Haven, CT 06520-8120}
\affiliation{Kavli Institute of Particle Astrophysics and Cosmology, Stanford University, 452 Lomita Mall, Stanford, CA 94305}

\author{C. M. Urry}
\affiliation{Yale Center for Astronomy and Astrophysics, and Physics Department, Yale University, PO Box 2018120, New Haven, CT 06520-8120}

\author{N. Cappelluti}
\affiliation{Physics Department, University of Miami, Coral Gables, FL 33155}

\author{J. T. Johnson}
\affiliation{Physics Department, University of Miami, Coral Gables, FL 33155}

\author{S. M. LaMassa}
\affiliation{Space Telescope Science Institute, 3700 San Martin Drive, Baltimore MD, 21218}

\author{T. T. Ananna}
\affiliation{Yale Center for Astronomy and Astrophysics, and Physics Department, Yale University, PO Box 2018120, New Haven, CT 06520-8120}
\affiliation{Department of Physics and Astronomy, Dartmouth College, 6127 Wilder Laboratory, Hanover, NH 03755}

\author{K. E. Kollmann}
\affiliation{Department of Physics, University of Maryland Baltimore County, Baltimore, MD 21250}

\begin{abstract}

The clustering of active galactic nuclei (AGN) sheds light on their typical large (Mpc-scale) environments, which can constrain the growth and evolution of supermassive black holes.
Here we measure the clustering of luminous X-ray-selected AGN in the Stripe 82X and XMM-XXL-north surveys around the peak epoch of black hole growth, in order to investigate the dependence of luminosity on large-scale AGN environment. 
We compute the auto-correlation function of AGN in two luminosity bins, $10^{43}\leq L_X<10^{44.5}$ \ergs\ at $z\sim 0.8$ and $L_X\geq 10^{44.5}$ \ergs\ at $z\sim 1.8$, and calculate the AGN bias taking into account the redshift distribution of the sources using three different methods. 
Our results show that while the less luminous sample has an inferred typical halo mass that is smaller than for the more luminous AGN, the host halo mass may be less dependent on luminosity than suggested in previous work.
Focusing on the luminous sample, we calculate a typical host halo mass of $\sim 10^{13}$ \Msun, which is similar to previous measurements of moderate-luminosity X-ray AGN and significantly larger than the values found for optical quasars of similar luminosities and redshifts. We suggest that the clustering differences between different AGN selection techniques are dominated by selection biases, and not due to a dependence on AGN luminosity. We discuss the limitations of inferring AGN triggering mechanisms from halo masses derived by large-scale bias.

\end{abstract}

\keywords{AGN}

\section{Introduction} \label{sec:intro}

The clustering statistics of active galactic nuclei (AGN) can provide insight into the relationship between accreting supermassive black holes and their host dark matter halos. 
By comparing the spatial distribution of an AGN sample to the well-understood clustering of halos, the typical AGN host halo mass can be inferred.
This allows for the characterization of AGN large-scale environments, which constrains the assembly and evolution of supermassive black holes.

Wide-area optical surveys such as the Sloan Digital Sky Survey (SDSS; \citealt{Petitjean:2018}) have detected tens of thousands of powerful quasars ($L_{\rm{bol}}> 10^{45}$ erg s$^{-1}$) across a wide range of redshifts. The resulting clustering amplitudes have constrained these quasars to reside in dark matter halos of a few $\times 10^{12}$ \Msun, largely independent of redshift \citep{Croom:2005,Coil:2007,Ross:2009,Shen:2009,White:2012,Eftekharzade:2015,Laurent:2017,He:2018,Timlin:2018}. This is consistent with what is expected for predominantly major merger-driven black hole accretion \citep{Hopkins:2008}, since galaxy major mergers are most probable in galaxy group environments.
However, popular scenarios of quasar/galaxy co-evolution predict an extended period of obscured black hole growth \citep[e.g.,][]{Hopkins:2006,Hickox:2009}, which is strongly selected against in optical surveys. The potentially large fraction of the luminous AGN missed in the optical limits the full picture. X-ray selection is a less biased AGN detection method, as high-energy photons can more easily penetrate the obscuring material and there is little contamination from the host galaxy. Wide-area, shallow surveys like \emph{Swift}/BAT and ROSAT have provided the host halo mass estimates for low-redshift obscured AGN \citep{Krumpe:2012,Krumpe:2017,Powell:2018}. But until recently, only deep pencil-beam X-ray surveys that detect low-to-moderate luminosity AGN (e.g., COSMOS; \citealt{Civano:2016,Marchesi:2016}) have been able to constrain the higher redshift environments closer to the peak of supermassive black hole accretion ($z\sim 1-3$; \citealt{Allevato:2011,Allevato:2014,Allevato:2016}). Previous clustering studies of moderate-luminosity X-ray AGN have found these AGN to reside in halos of $\sim 10^{13}$ \Msun\ up to $z\sim2$, statistically higher masses than found for optical quasars \citep[see also,][]{Allevato:2011,Starikova:2011,Cappelluti:2012,Mountrichas:2012}. It remains to be seen whether this difference is because of a luminosity dependence in AGN clustering statistics (due to disparate triggering processes), or because of biases resulting from the different AGN selection methods \citep{Mendez:2016}.

In this study, we combine two of the largest-area deep X-ray surveys to probe the environments of the most luminous X-ray-selected AGN. The Stripe 82X \citep{Lamassa:2013B,Lamassa:2016} and XMM-XXL-north \citep{Pierre:2016} surveys have a combined area of $\sim 38$ deg$^2$, detecting AGN radiating up to $L_{bol}\sim 10^{47}$ \ergs\ at redshifts $z\sim 1-3$. This fills the missing tier between the wide/shallow X-ray surveys like BASS \citep{Koss:2017} and the deep pencil-beam X-ray surveys like COSMOS, and provides a link between the X-ray AGN and optically-selected quasars with similar luminosities and redshifts. Defining two bins of luminosity, we compare the derived halo masses of each AGN subsample with previous studies in the literature in order to investigate the luminosity dependence of AGN clustering. Throughout this paper, we assume Planck 2015 cosmology (\citealt{Planck:2015}; $H_0=100h$ km/s/Mpc, $h=0.677$, $\Omega_{m,0}=0.307$, $\Omega_{b,0}=0.0486$).

\section{Data} \label{sec:data}
\subsection{Stripe 82X}
The Stripe 82 X-ray survey (S82X) comprises several fields of X-ray coverage in the Sloan Digital Sky Survey (SDSS) Stripe 82 Legacy field. This includes three regions observed with {\it XMM-Newton} observations in cycles 10 and 13; two 2.3 deg$^2$ patches (AO10) and one 15.6 deg$^2$ patch (AO13) of contiguous area. The details of the data analysis are given by \cite{Lamass:2013A,Lamassa:2013B,Lamassa:2016}. The area covered as a function of the flux limit is shown in Fig. \ref{fig:AF}. In addition to X-ray coverage, there is an abundance of multiwavelength data in this field spanning the entire electromagnetic spectrum: ultraviolet (GALEX), optical (SDSS), near-infrared (VHS, UKIRT, WISE), mid-infrared (WISE, Spitzer), far-infrared (Herschel), millimeter (ACT), and radio (FIRST, VLA). The counterparts were matched using a Maximum Likelihood Estimator and unique identifications were verified by eye \citep{Ananna:2017}.

At present, $54\%$ of the X-ray sources have spectroscopic redshifts, obtained both from publicly available catalogs \citep{Strauss:2002,Jones:2004,Garilli:2008,Croom:2009,Drinkwater:2010,Coil:2011,Ahn:2012,Newman:2013,Alam:2015}, follow-up programs at facilities on Palomar, WIYN, and Keck by our team \citep{Lamassa:2016}, and through a dedicated SDSS-IV eBOSS follow-up survey \citep{Lamassa:2019}. For objects without spectroscopy, high-quality photometric redshifts have been calculated from the multi-epoch photometry ($\sigma_z =0.06$, with an outlier fraction of 13.7\%) using the \texttt{LePhare} software \citep{Arnouts:1999,Ilbert:2006}, as discussed in detail in \cite{Ananna:2017}.

For the fraction of S82X AGN with spectroscopic redshifts, we used the publicly available {\tt Cigale} code \citep{Burgarella:2005,Noll:2009,Serra:2011,Ciesla:2015,Boquien:2019} to fit the full spectral energy distributions and estimate the host galaxy stellar masses. We assumed \citealt{Maraston:2005} stellar population libraries with a \cite{Salpeter:1955} IMF, and used the \cite{Calzetti:2000} dust attenuation law. The \cite{Fritz:2006} templates were used to model the AGN component. More details are located in the Appendix. This resulted in stellar masses estimates and their uncertainties for 2757 total AGN.

We selected AGN in the AO10 and AO13 regions of Stripe82X with \emph{det\_ml} $>15$, corresponding to being detected with a significance over $5\sigma$ (\emph{det\_ml}$ \equiv -\ln$ P, where P is the Poissonian probability that the detection is due to a random background fluctuation). We further selected the sources that have either a spectroscopic redshift or a firm photo-z, defined as the integrated probability within $\pm 1\sigma$ of the best$-$fit redshift exceeding 
$90\%$ (i.e., `PDZ\_BEST' $>90$; see \citealt{Ananna:2017}). We utilized the full photo-z probability distribution functions in our clustering analysis (see Section 3.2). This `PDZ\_BEST' threshold was chosen empirically to minimize the uncertainty on the measurement, balancing the inclusion of more photo-z objects against smoothing out the line-of-sight clustering signal. There are 2337 total AGN meeting these criteria (344 with photo-z's only).

\subsection{XMM-XXL}
This work uses the XMM-XXL catalogue presented by \cite{Liu:2016} based on the X-ray reduction pipeline described by \cite{Georgakakis:2011}. The XMM-XXL-north field is an $\sim$18 deg$^{2}$ region observed by \emph{XMM-Newton} \citep{Pierre:2016} with overlapping spectroscopic coverage from the BOSS program \citep{Alam:2015}. X-ray detections are defined as having \emph{det\_ml} $>12.42$, and 2578 of those have optical classifications and reliable spectroscopic redshift measurements \citep[33\%;][]{Menzel:2016}. In addition, X-ray spectral analysis has been performed to obtain column densities for each AGN, as detailed in \cite{Liu:2016}. 

\begin{figure}
\centering
\includegraphics[width=.5\textwidth]{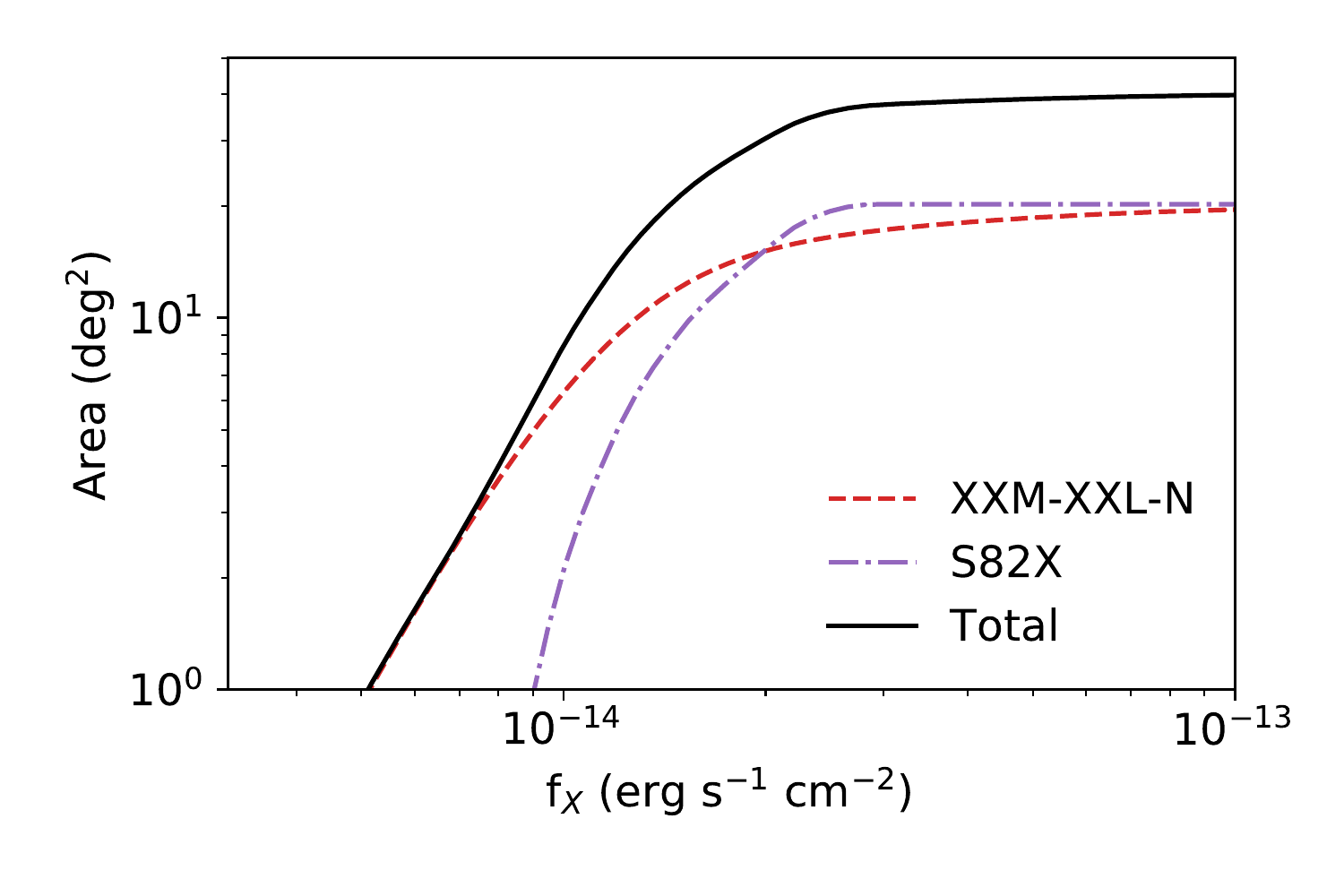}
\caption{Full-band (0.5-10 keV) sensitivity curves for S82X (purple, from \citealt{Lamassa:2016}, detection threshold \emph{det\_ml} $>15$), XMM-XXL-north (red, from \citealt{Liu:2016}, detection threshold \emph{det\_ml} $>12.42$), and combined (black).}
\label{fig:AF}
\end{figure}

We selected the AGN in XMM-XXL-north with spectroscopic redshifts in the DR12 BOSS footprint. While the incompleteness of the spectroscopic redshifts affects the clustering on small angular scales ($<0.03$ deg; \citealt{Mountrichas:2016}), this effect is small for the projected scales that we are interested in at the effective redshifts of our samples ($z\sim 0.7$ and $z\sim 1.8$, corresponding to 0.9 and 1.2 Mpc h$^{-1}$, respectively). The integrated sensitivity curves are shown in Figure \ref{fig:AF} \citep{Georgakakis:2008}.

\begin{table}
\begin{center}
\begin{tabular}{ c c c c c c }
\hline
& Field & Area & $N$ & $\langle z \rangle$ & $\log \langle L_X \rangle$ \\ 
& &  (deg$^{2}$) &  & & (\ergs)\\ 
\hline
\hline
High$-L$ & S82X-AO10 & 4.6 & 169 & 1.81 & 45.06\\ 
& S82X-AO13 & 15.6 & 732 & 1.75 & 45.04\\  
& XMM-XXL-N & 18.1 & 1003 & 1.84 &  45.03\\
& Total & 38.3 & 1904 & 1.80 & 45.04\\
 \hline
Low$-L$ & S82X-AO10 & 4.6 & 236 & 0.72 & 44.06 \\ 
& S82X-AO13 & 15.6 & 967 & 0.76 &  44.06 \\  
& XMM-XXL-N & 18.1 & 1137 & 0.84 & 44.06 \\
& Total & 38.3 & 2335 & 0.80 & 44.06\\
 \hline
\end{tabular}
\caption{Characteristics of the AGN samples used in this work for each contiguous field, including the S82X areas (AO10 and AO13) and the XMM-XXL-north area. The high$-L$ sample is defined as $\log~L_{0.5-10~keV}\geq44.5$ [\ergs] and the low$-L$ bin is $43\leq \log~L_{0.5-10~keV}<44.5$ [\ergs].} 
\end{center}
\label{tb:t1}
\end{table}

\subsection{Luminosity Selection}
The observed full-band X-ray luminosities ($0.5-10$ keV) were calculated for the AGN from fluxes ($f_X$) in both surveys \citep{Lamassa:2016,Liu:2016} via $L_X=4\pi d_{L}^{2} f_X$, where $d_L$ is the luminosity distance. $\Gamma=2$ was assumed for the $k-$correction, which is the median spectral index of the XXL AGN \citep{Liu:2016}. This does not change the fluxes since the correction scales as $(1+z)^{\Gamma - 2}$. While column densities have been measured for the XMM-XXL-north sample from their X-ray spectral fitting, this is still in progress for the S82X sample. We verified that our calculated luminosities were similar to the rest-frame intrinsic luminosities measured in \cite{Liu:2016}. We defined our high-luminosity bin as AGN with $\log L_X$[\ergs]$> 44.5$, and our lower-luminosity bin as $43<\log L_X$[\ergs]$< 44.5$. 

To see whether obscuration could significantly change the sample by underestimating the intrinsic luminosities, we estimated how many additional AGN would be included in our high$-L$ selection by assuming an $N_H$ distribution matching the XXL AGN. After estimating the correction to the observed $L_X$, we find that only $\lesssim 4\%$ of the sample would change.

\begin{figure}
\centering
\includegraphics[width=.49\textwidth]{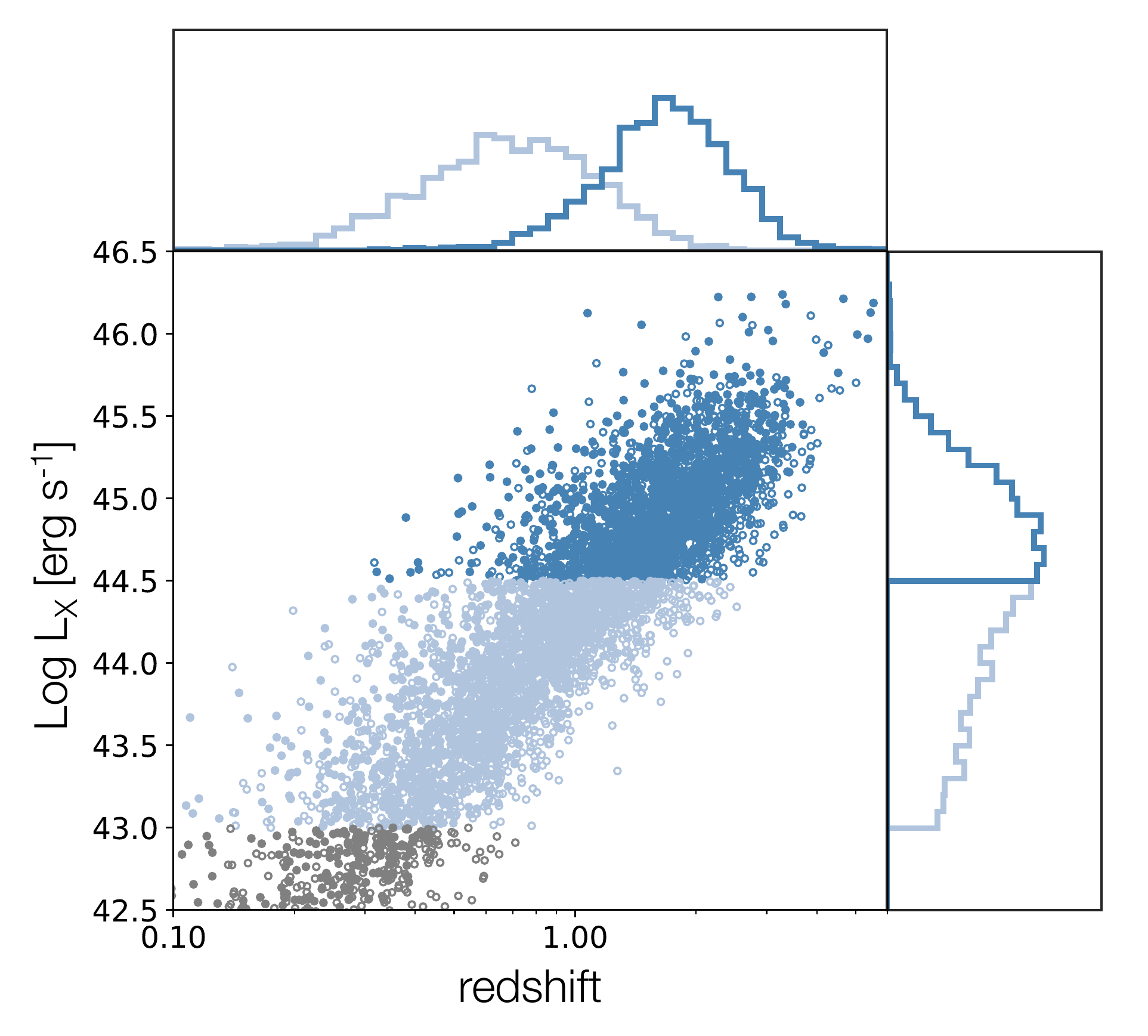}
\caption{Observed 0.5-10 keV X-ray luminosity vs. redshift for our combined sample of AGN from the Stripe 82X (filled circles) and XMM-XXL-North (open circles) surveys. The dark blue points correspond to the high$-L$ bin and the light blue points correspond to low$-L$ bin.}
\label{fig:LvZ}
\end{figure}

\begin{figure*}
\centering
\includegraphics[width=.75\textwidth]{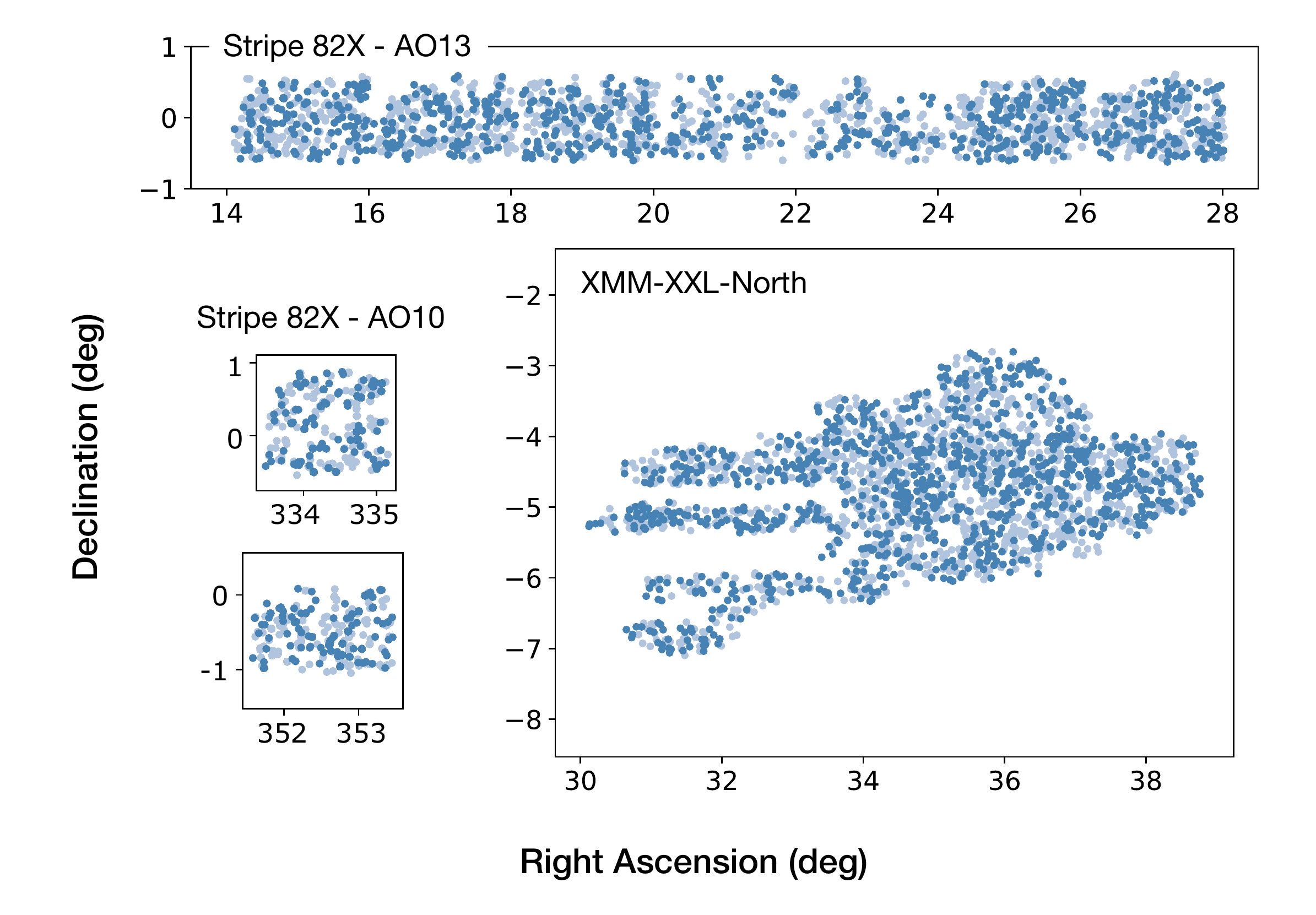}
\caption{Positions of the high$-L$ (blue) and low$-L$ (light blue) AGN samples used in this work, including those from the contiguous A010 and A013 regions of S82X (left and top) and those from the XMM-XXL-north field (right).}
\label{fig:radec}
\end{figure*}

The redshifts vs. luminosities of the AGN are shown in Figure \ref{fig:LvZ}, and their spatial coordinates are shown in Figure \ref{fig:radec} for each field. The weighted average redshift (including the full photo$-z$ PDFs; see Section 3.2) of our high$-L$ (low$-L$) bin is 1.80 (0.80), with a weighted average $L_X$ of $\sim 10^{45}$ ($\sim 10^{44}$) erg s$^{-1}$. The characteristics of the catalog disaggregated by field are given in Table \ref{tb:t1}. Note that for some objects on the luminosity thresholds, only parts of their photo$-z$ probability distribution functions were used (only the parts that satisfy the luminosity requirements based on the redshifts and flux of the object). Although we count each such object as 1 in the numbers given in Table \ref{tb:t1}, they count as fractional objects in the clustering analysis.

\section{Clustering Methodology} \label{sec:methods}

The spatial 2-pt correlation function $\xi(r)$ quantifies the excess probability that a pair of objects are separated by distance $r$. A larger amplitude of $\xi$ corresponds to a more clustered sample, while $\xi=0$ suggests that it is randomly distributed in space. 

The galaxy correlation function is a superposition of two terms; the 1-halo term, which dominates on scales $\lesssim 1$ \Mpch\ and measures the clustering of galaxies within the same dark matter halo, and the 2-halo term, which dominates on scales $> 1$ \Mpch\ and measures the clustering of galaxies in distinct dark matter halos. The amplitude of the latter gives an estimate of the host halo mass of the sample (see Section \ref{sec:halomass}).

We use the Davis \& Peebles estimator \citep{Peebles:1983} to compute the weighted correlation function in bins perpendicular ($r_p$) and parallel ($\pi$) to the line of sight:

\begin{equation}
\xi(r_{p},\pi) = \frac{AA}{AR} - 1 
~~,
\end{equation}

\noindent where AA and AR are the weighted numbers of AGN-AGN and AGN-random pairs in a bin of $r_{p}$ and $\pi$, respectively: 

\begin{equation}
AA = \sum_{i,j}\frac{\omega_{i}\times \omega_{j}}{W^2}
~~,
\end{equation}

\begin{equation}
AR = \sum_{i,j}\frac{\omega_{i}}{W\times N_{random}}
~~.
\end{equation}

\noindent $\omega$ refers to the redshift weight assigned to each AGN (see section 3.2) and $W$ is the total sum of the weights. Indices are summed over all AGN pairs. We use the {\tt CorrFunc} software for the weighted pair counting \citep{Sinha:2017}.

To eliminate redshift-space distortions, we integrate the $\pi$ dimension to obtain the projected correlation function $w_{p}$:

\begin{equation}
w_{p} = 2 \int^{\pi_{\rm{max}}}_{0} \xi(r_{p},\pi) d\pi
~~.
\end{equation}

\noindent $\pi_{\mathrm{max}}$ is chosen as the value in which the 2-halo term of $w_p$, averaged over scales from $1-10$ \Mpch, converges and only gets noisier for any higher value. We empirically determined this value to be roughly 60 \Mpch. This is large enough to integrate over the higher average redshift smearing caused by including $\sim 10\%$ photo$-z$ sources.

\subsection{Random Catalog Generation}

We constructed four random AGN catalogs, one for each contiguous field in our sample. Working with each field separately, we first smoothed the redshift distribution of the data by a Gaussian kernel with $\sigma_{z}=0.2$, and chose a redshift for each random AGN by drawing from the smoothed distribution. The smoothing scale $\sigma_{z}$ corresponds to scales $\gtrsim 100$ \Mpch\ throughout our redshift range, so that large-scale structures are smoothed over and not reflected in the redshift distribution of the random catalog. 

For the angular coordinates of the random sample, the sensitivity maps of each survey were utilized. For XMM-XXL-north, we used the available sensitivity map that was constructed from the method described in \cite{Georgakakis:2008}. We produced the S82X sensitivity maps by the same method from the survey's background and exposure maps,
calculating the limiting flux for a $5.1\sigma$ detection in bins of size $32^{\prime\prime}\times 32^{\prime\prime}$.

We first randomized RA and DEC for the random catalog in the footprints of each field, and then assigned each a flux drawn from the Log $N-$Log $S$ distribution from \cite{Lamassa:2016}. Derived from simulations using fits to deeper data, the Log $N-$Log $S$ distribution describes the number counts of the data folded into the survey's area-flux curve. We then kept the random sources whose flux values were larger than the sensitivity at their respective positions. 

With the resulting fluxes and redshifts of the randoms, we then imposed the same luminosity limits for each defined luminosity bin and downsampled the catalogs to ensure that the overall redshift distributions of each luminosity bin matched that of the smoothed distributions of the data.  We verified that the final flux distributions, as well as the relations between redshift and luminosity, were similar between the randoms and data for each bin. The resulting catalogs were constructed to be $\sim 100$ times larger than the data catalogs in order to minimize Poisson noise.

\subsection{Utilizing Full Photo-z PDFs}

While 54\% of the sources in S82X currently have spectroscopic redshifts, nearly all remaining objects have photometric redshifts \citep{Ananna:2017}. In order to maximize the information we extract from those photometric redshifts, we utilize the full PDFs following the method in \cite{Allevato:2016}; each galaxy is essentially `spread out' through redshift space and sampled by its normalized PDF. One AGN therefore becomes many (depending on the PDF sampling) with associated redshift weights, which equal the PDF value at their redshift. Each AGN photo$-z$ PDF is normalized such that $\sum_i \mathrm{PDF}(z_i)=1$.

\subsection{Error Estimation}
The correlation function uncertainties were estimated via the jackknife re-sampling technique. We divided the AGN sample into 25 patches on the sky \citep[e.g.,][]{Powell:2018}, each containing $2-6\%$ of the data, and repeated the measurement when excluding each patch ($w_k$). The scales of the patches are larger than the scales of the 2-halo term at these redshifts, and so each patch is assumed to be independent. The covariance matrix is estimated by:

\begin{equation}
\begin{split}
C_{i,j} = \frac{M}{M-1} \sum_{k}^{M} \Big[w_{p,k}(r_{p,i}) - \langle w_{p}(r_{p,i})\rangle\Big]\\
\times\Big[w_{p,k}(r_{p,j}) - \langle w_{p}(r_{p,j})\rangle \Big] ~~,
\end{split}
\end{equation}

\noindent where $M$ is the number of jackknife samples (25). The errors on $w_p$ for each $r_p$ bin are the square roots of the diagonals:  $\sigma_i=\sqrt{C_{i,i}}$.

\section{Halo Mass Estimation} \label{sec:halomass}

In the standard halo model approach, galaxies reside in dark matter halos that have collapsed and virialized at the peaks of the underlying dark matter distribution. The galaxy halo occupation distribution (HOD) refers to the probability $P(N|M_h)$ that $N$ galaxies reside in a halo with mass $M_h$ \citep[e.g.,][]{Cooray:2002}. The clustering statistics of galaxies therefore depend only on cosmology (governing how halos cluster) and the associated HOD. On scales greater than $\sim 1$ \Mpch, the clustering of galaxies in separate halos dominates the correlation function (the 2-halo term). The amplitude of this term relative to that of dark matter halos, defined as the bias, can estimate the typical halo mass that the sample resides in \citep[e.g.,][]{Tinker:2010}. 

The AGN bias is estimated by taking the ratio between the 2-halo terms of the AGN and dark matter correlation functions, averaged over scales from $1-10$ \Mpch:

\begin{equation}
b_{AGN} = \sqrt{\frac{w_{p,AGN}}{w_{p,DM}}}
~~.
\end{equation}

The projected dark matter correlation function, $w_{p,DM}$, is calculated from integrating the real space correlation function, obtained by \citep{Peebles:1983}:

\begin{equation}
w_{p,DM}(r_p) = 2\int_{r_p}^{r_{\mathrm{max}}} \frac{\xi_{DM}(r) rdr}{\sqrt{r^2-r_p^2}}
~~,
\end{equation}

\noindent where $r_{\mathrm{max}}=\sqrt{\pi_{\mathrm{max}}^2 + r_p^2}$ and $\xi_{DM}$ is the Fourier transform of the matter power spectrum $P(k)$:

\begin{equation}
\xi_{DM}(r) = \frac{1}{2\pi^2}\int_{r_p}^{\infty} P(k)k^2 \left(\frac{\sin(kr)}{kr} \right)dk
~~.
\end{equation}

\noindent $P(k)$ is calculated assuming a spectral index $n=1$ with the transfer function from \cite{Eisenstein:1998}, using the publicly available {\texttt{hmf}} software \citep{Murray:2013,hmf:2014}. This software also includes the nonlinear corrections to the power spectrum from \cite{Smith:2003} and \cite{Takahashi:2012}.

We derive the typical halo mass of the AGN in our sample using three methods: (1) we calculate the AGN bias at the weighted average redshift of the sample; (2) similar to (1) but at the effective redshift of the sample; and (3) we take the full redshift distribution into account and compute the weighted bias. We describe each method below.

For methods (1) and (2), the typical halo mass is inferred from the AGN bias using the analytic halo bias function $b_{\mathrm{T}10}(\nu)$ from \cite{Tinker:2010} with $\delta_{\mathrm{halo}}=200$, where $\nu=\delta_c/\sigma(M)$. The quantity $\sigma(M)$ is the root-mean square of mass density fluctuations within a sphere containing mass $M$, given by:

\begin{equation}
\sigma^2(M) = \int P(k,z)\hat{W}(k,R)k^2dk
\end{equation}

We use a spherical top-hat window function for $\hat{W}(k,R)$, and $R=(3M_h/4\pi \bar{\rho})^{1/3}$ is the radius enclosing mass $M$, where $\bar{\rho}$ is mean density of the universe. The typical AGN host halo mass is the value that satisfies

\begin{equation}
b_{AGN} = b_{\mathrm{T}10}(M,z)
~~,
\end{equation}

\noindent where $z$ is either the weighted average ($\langle z\rangle $) or effective redshift ($z_{eff}$) of the sample. These are defined as the following:

\begin{equation}
\langle {z} \rangle= \frac{\sum_{i} \omega_i ~z_{i}}{\sum_{i} \omega_i} 
~~,
\end{equation}

\begin{equation}
z_{eff} = \frac{\sum_{i,j} \omega_i ~\omega_j ~z_{pair}}{\sum_{i,j} \omega_i~ \omega_j}
~~,
\end{equation}

\noindent where $z_{pair}=(z_i+z_j)/2$ and $i$ and $j$ sum over the AGN pairs.

For the third method, we take into account the full redshift range of the sample as well as the growth of structure throughout that range, following the method in \cite{Allevato:2011} to compute the weighted AGN bias. This method assumes that the HOD of the AGN is a delta function, such that all AGN reside in halos of a given mass. While this is not physical, the halo mass obtained from this method would be comparable to the average host halo mass of the sample assuming a somewhat narrow distribution of host halo masses. This is expected if major mergers predominantly trigger AGN, as major mergers are most efficient in group environments. The limited range of luminosity of our sample may also satisfy this assumption, and the little evolution of host halo masses with redshift seen in previous studies is consistent with this. However, we discuss the limitations and caveats of this assumption in Section 6. Nevertheless, this method is a good test to see whether methods (1) or (2) can be valid for samples spanning a broad range of redshifts.

Each pair is weighted by the bias, $b_i = b_{\mathrm{T}10}(M,z_i)$, and growth factor ($g_i$) at its redshift. The AGN pairs are then summed over and normalized:
 
\begin{equation}
\bar{b}(M) = \sqrt{\frac{\sum_{i,j} b_i(M)~ b_j(M)~ g_i~ g_j~\omega_i~ \omega_j}{W_{AGN}^{2}}}~~,
\end{equation}

\begin{equation}
\bar{z}(M) = \frac{\sum_{i,j} b_i(M) ~b_j(M) ~g_i~ g_j~\omega_i~ \omega_j ~z_{pair}}{\sum_{i,j} b_i(M)~ b_j(M)~ g_i~ g_j~\omega_i~ \omega_j}
~~.
\end{equation}

\noindent The AGN host halo mass $M_{h, AGN}$ is then the value that satisfies

\begin{equation}
\bar{b}(M_{h, AGN}) = \sqrt{\frac{w_{p,AGN}}{w_{p,DM}(z=0)}}
~~.
\end{equation}

The AGN bias ($b_{AGN}$) quoted is then the value $b_{\mathrm{T}10}(M_{h, AGN},\bar{z})$ for the resulting halo mass and weighted redshift calculated.

\subsection{Modeling the predictions for similar inactive galaxies}
In this section we describe the process of utilizing the stellar mass estimates of the AGN host galaxies in our sample to compute the predicted clustering based on this property alone. 
This was done via the approach presented in \cite{Powell:2018}, in which we populate dark matter halos from $N-$body simulation snapshots with {\texttt{halotools}} \citep{Hearin:2017} and forward model the stellar mass incompleteness to match our data selection.

We used snapshot Rockstar halo catalogs from the Consuelo simulation \citep{Behroozi:2013B,Behroozi:2013C} near the effective redshifts of each AGN bin ($z=0.65$ and $z=1.77$). The Consuelo simulation has a simulation box size of 420 \Mpch, and a particle resolution of $\sim 10^9$ \Msun\ (complete for halos $\gtrsim 10^{11}$ \Msun). At the center of each halo and subhalo, we placed a mock galaxy. Using the stellar mass-halo mass relation from \cite{Behroozi:2010}, we assigned each mock a stellar mass, and then subsampled the full mock catalog such that its stellar mass distribution matched that of our AGN. 

The stellar mass distributions of each luminosity bin of our AGN hosts were obtained from the \texttt{Cigale} estimates of the S82X spectroscopic sample. Two subsamples of 500 AGN were chosen from this stellar mass catalog that satisfied the same luminosity thresholds and had the same redshift distributions as each of our luminosity bins. The stellar mass distributions of the 500 AGN were then assumed to represent that of the full Stripe82X+XMM-XXL-north samples.

The averaged scale-dependent clustering of 20 mock realizations was then compared to the AGN clustering results. This checked for consistency with the prediction for inactive galaxies, where stellar mass primarily drives the clustering statistics. Uncertainties on this prediction were obtained by assuming $\pm 0.25$ dex offsets of the stellar mass distributions. The magnitude of the offsets represent typical errors on the estimates according to {\tt Cigale}.

\section{Results}
\label{sec:results}

\begin{figure}
\centering
\includegraphics[width=.5\textwidth]{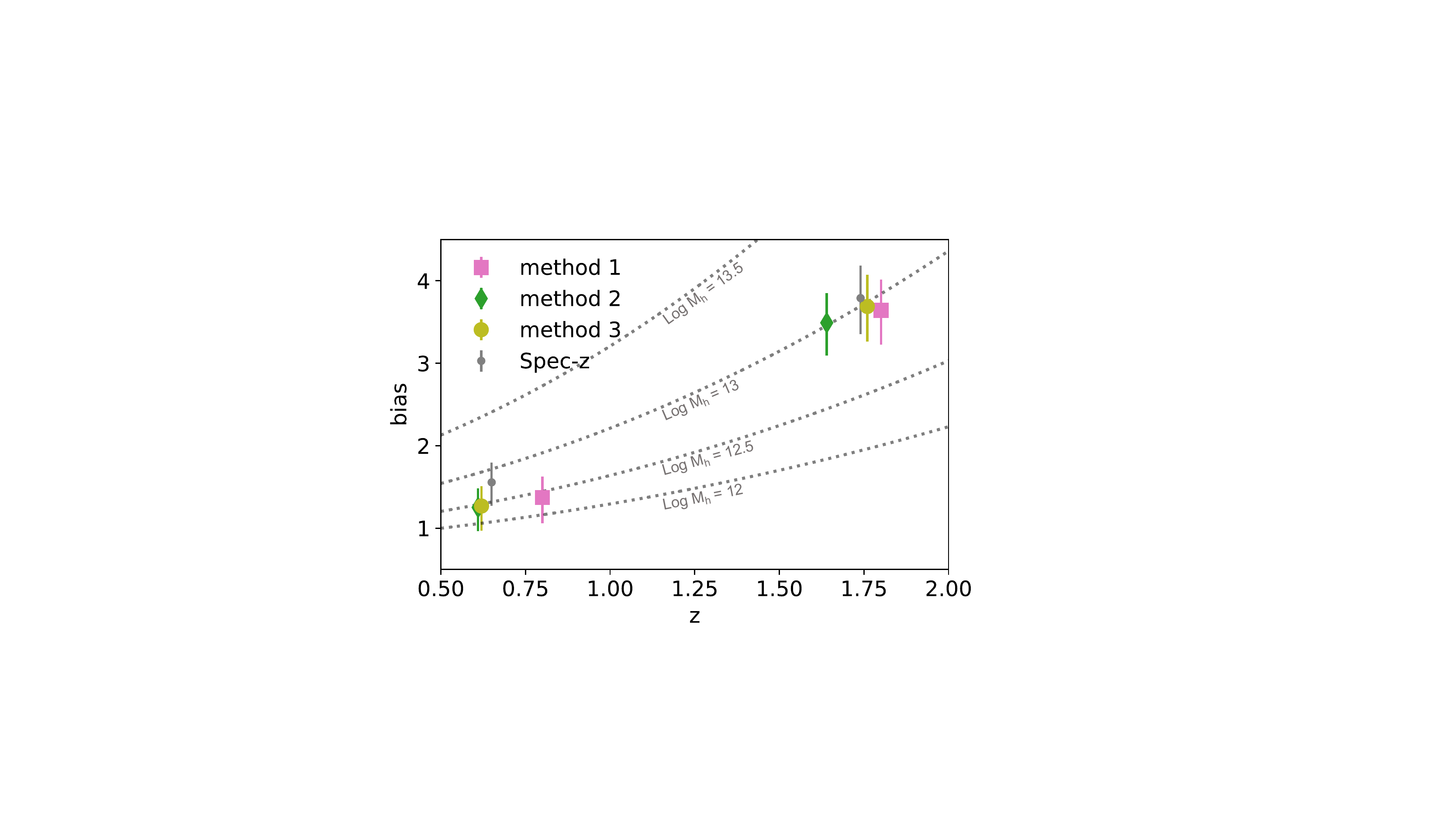}
\caption{Bias values for our AGN sample calculated by the three different methods described in the text (colored points). The biases calculated by method 3 using only AGN with spectroscopic redshifts are shown in gray. The various methods are consistent with each other. Lines of constant halo mass are shown for reference by the labeled dotted lines (in units of \Msun).}
\label{fig:mh}
\end{figure}

 \begin{table}
\begin{center}
\begin{tabular}{ c c c c c  }
\hline
& Method & bias & redshift & $\log M_h$ \\ 
& &  &  & [\Msun]\\ 
\hline
\hline
High-L & 1 & $3.64^{+0.37}_{-0.42}$ & 1.80 & $12.93^{+0.13}_{-0.17}$ \\ 
& 2 & $3.49^{+0.36}_{-0.40}$ & 1.64 & $13.01^{+0.13}_{-0.16}$ \\  
& 3 & $3.69^{+0.38}_{-0.43}$ & 1.76 & $12.98^{+0.13}_{-0.17}$ \\
& Spec-z only & $3.79^{+0.39}_{-0.44}$ & 1.74 & $13.03^{+0.13}_{-0.16}$ \\
 \hline
Low-L & 1 & $1.37^{+0.25}_{-0.31}$ & 0.80 & $12.39^{+0.33}_{-0.65}$ \\ 
& 2 & $1.25^{+0.23}_{-0.29}$ & 0.61 & $12.44^{+0.35}_{-0.71}$ \\  
& 3 & $1.27^{+0.24}_{-0.30}$ & 0.62 & $12.42^{+0.35}_{-0.74}$ \\
& Spec-z only & $1.56^{+0.24}_{-0.28}$ & 0.65 & $12.83^{+0.25}_{-0.41}$\\
 \hline
\end{tabular}
\caption{Bias and halo mass measurements of X-ray-selected quasars, using the 3 different methods described in the text. Also shown are the results when using the sample with spectroscopic redshifts only (using method 3).} 
\end{center}
\label{tb:t2}
\end{table}

\begin{figure*}
\centering
\includegraphics[width=.45\textwidth]{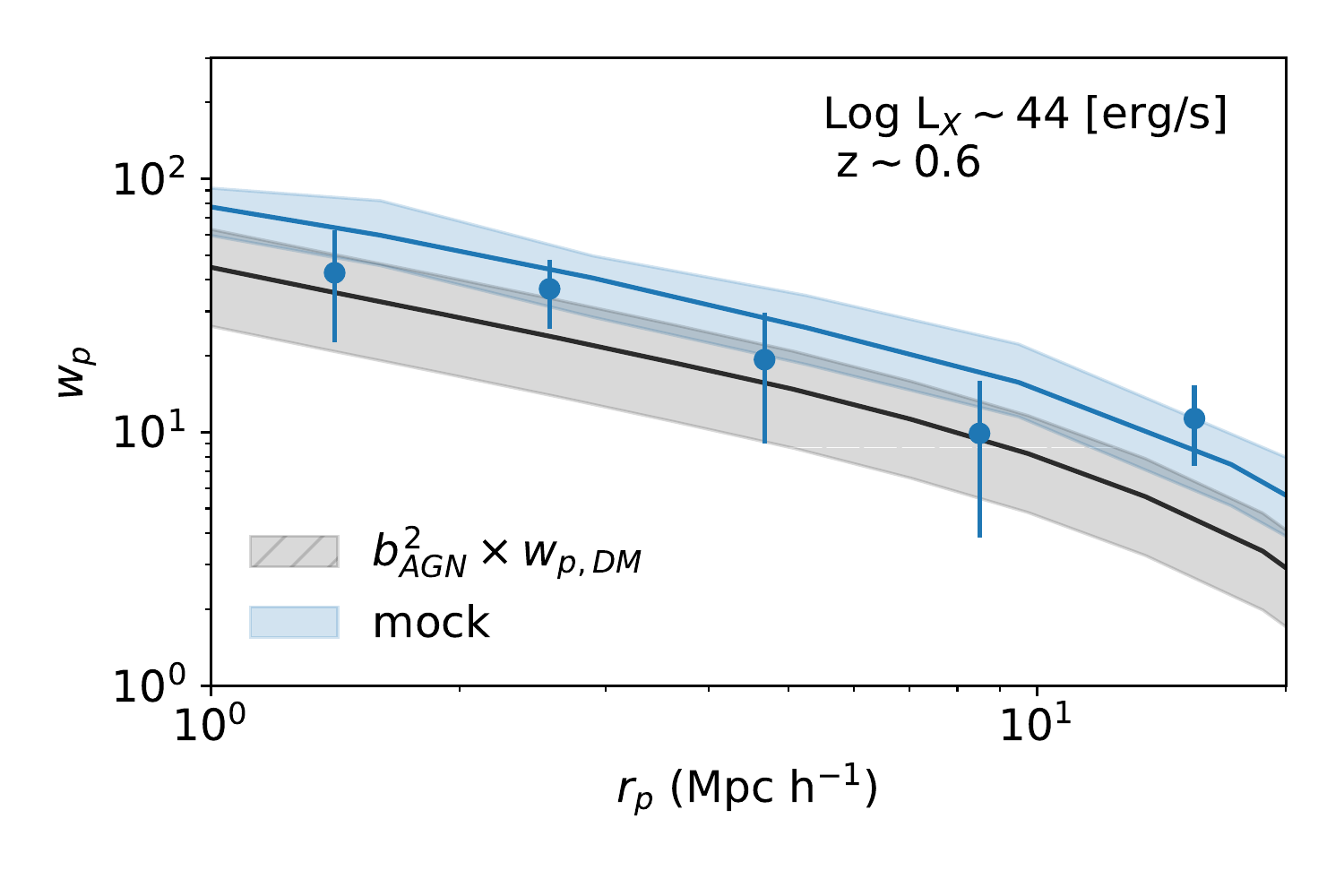}
\includegraphics[width=.45\textwidth]{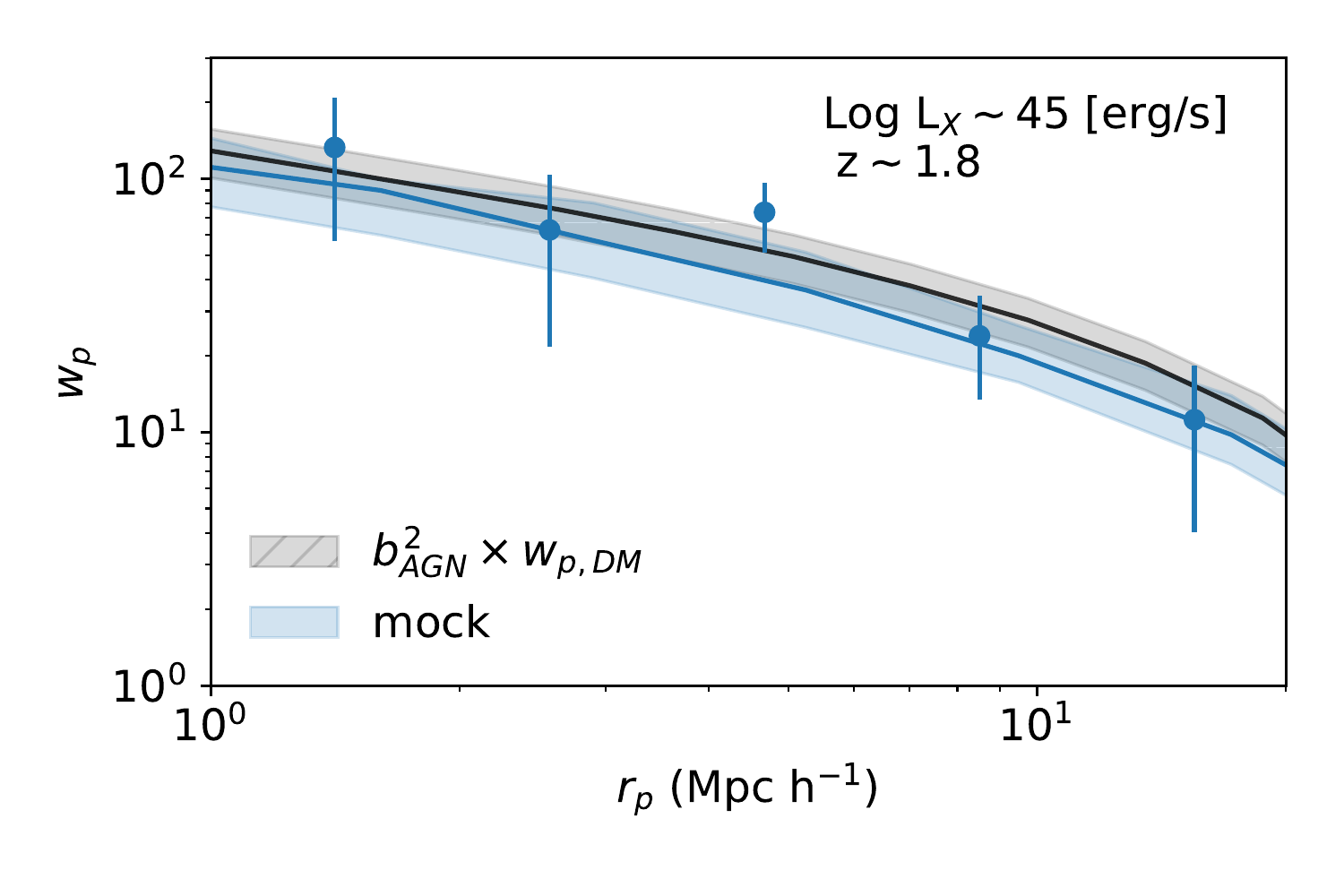}
\caption{Projected correlation functions of the high-luminosity AGN bin ({\it left}) and lower-luminosity AGN bin ({\it right}) with associated models. 
The black lines show $b_{AGN}^{2}\times w_{p,DM}$ using the bias values calculated via method 3, and the shaded regions correspond to the one-sigma uncertainties on the bias measurements. The projected correlation functions of the mock samples that have the same stellar mass distributions of the AGN are shown by the blue lines. The bounds of the shaded blue areas assume $\pm 0.25$ dex offsets of stellar mass distributions.}
\label{fig:wp}
\end{figure*}

The bias values and corresponding assumed redshifts from each method are shown in Figure \ref{fig:mh} for both luminosity bins. The luminous X-ray-selected quasars at $z\sim 1.8$ were calculated to have biases of $3.64^{+0.37}_{-0.42}$, $3.49^{+0.36}_{-0.40}$, and $3.69^{+0.38}_{-0.43}$ for methods 1, 2, and 3, respectively. This corresponds to halo masses of $12.93^{+0.13}_{-0.17}$, $13.01^{+0.13}_{-0.16}$, and $12.98^{+0.13}_{-0.17}$ in log units of \Msun. For the lower-luminosity bin at $z\sim 0.7$, biases of $1.37^{+0.25}_{-0.31}$, $1.25^{+0.23}_{-0.29}$, and $1.27^{+0.24}_{-0.30}$ were found, which corresponds to halo masses of $12.39^{+0.33}_{-0.65}$, $12.44^{+0.35}_{-0.71}$ and $12.42^{+0.35}_{-0.74}$ in log units of \Msun. Consistent halo masses are found when only using AGN with spectroscopic redshifts via method 3, verifying that the use of the photo$-z$ distribution functions did not significantly shift the measurements. However, it should be noted that the spectroscopic sample is biased toward including brighter, unobscured objects, and so an exact match in halo mass is not expected. The results are summarized in Table \ref{tb:t2}. 

The three methods used to calculate the typical halo masses are consistent with each other. This indicates that using the median redshift of a sample with a broad redshift range for the halo mass calculation does not systematically skew results {\it when assuming a somewhat narrow distribution of host halo masses across the entire redshift range}. While this assumption may not be valid (the implications of which are discussed in the following section), it allows us to compare with previous measurements from the literature that have used  narrower redshift ranges and assumed one redshift value (as opposed to the full distribution) for their bias calculations.

The luminous quasars were calculated to reside in halos of $\sim 10^{13}$ \Msun, slightly higher than the value found for the lower-luminosity/lower-redshift sample ($3\times 10^{12}$ \Msun). We note that the value for the lower-luminosity bin is also consistent with what was reported for the overlapping AGN in the redshift range $0.5<z<1.2$ from XMM-XXL-north field alone, which was calculated via cross-correlating with galaxies for improved statistics \citep{Mountrichas:2016}. When taking each field separately, the variance between the fields were within error of each other.

Figure \ref{fig:wp} shows the measured projected correlation functions of both luminosity samples in several bins of $r_p$, with the resulting scale-dependent linear bias models. The models were calculated via $b_{AGN}^{2}\times w_{p,DM}(\bar{z})$ using the bias and redshift values obtained from our third method. Also shown are the correlation function predictions from the generated stellar mass-matched mock samples. We find consistency with the prediction based on stellar mass alone for both AGN samples, although the prediction is marginally higher for the lower luminosity bin (left-hand panel). More data is needed to determine whether or not this becomes a significant difference.

\section{Discussion}
\label{sec:disc}

\begin{figure*}
\centering
\includegraphics[width=\textwidth]{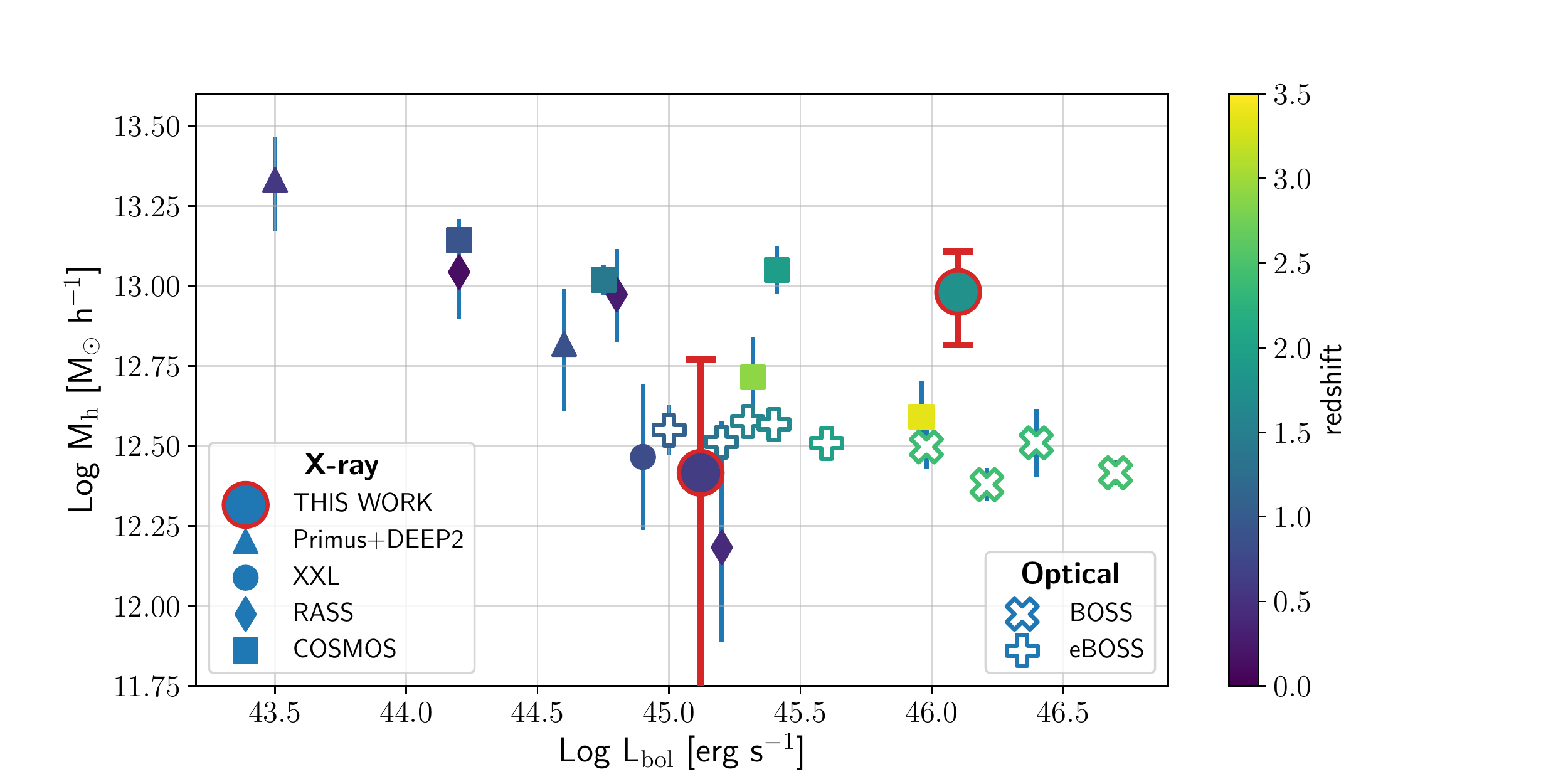}
\caption{AGN host halo mass as a function of bolometric luminosity for X-ray (filled squares, COSMOS, \citealt{Allevato:2011,Allevato:2014,Allevato:2016}; filled triangles, Primus fields, \citealt{Mendez:2016}; filled diamonds, RASS, \citealt{Krumpe:2012}; filled circles, XMM-XXL, \citealt{Mountrichas:2016}) and optical (x's, BOSS, \citealt{White:2012,Eftekharzade:2015}; thin crosses, eBOSS \citealt{Laurent:2017}) AGN. The color of the data points correspond to the effective redshift of the sample (right-hand colorbar). The results from this work are shown by the red-outlined circles.}
\label{fig:lit}
\end{figure*}

The halo masses calculated for each luminosity/redshift bin of our sample agrees with previous studies that found marginal or insignificant luminosity and/or redshift dependencies of the AGN clustering amplitude within the same sample of objects \citep[e.g.,][]{Ross:2009,Allevato:2011,Eftekharzade:2015,Starikova:2011,Laurent:2017}. While the halo masses of each bin differ by $\sim 0.5$ \Msun\, the large errors on the low-L measurement render this difference uncertain. Additionally,
the dependencies on luminosity that have been previously found within a survey typically go in the other direction, where the higher luminosity objects have smaller halo masses \citep{Krumpe:2012,Allevato:2011,Allevato:2014,Allevato:2016,Mendez:2016}. There are similar differences between {\it separate} surveys as well \citep{Mountrichas:2016}, including the typical disparity of halo masses found for moderate-luminosity X-ray AGN and luminous quasars at moderate redshifts \citep[e.g.,][]{Cappelluti:2012}.

Figure \ref{fig:lit} shows the comparison of our measurement with recent, previous projected clustering measurements from the literature for both X-ray and optical samples, as a function of average bolometric luminosity. The halo masses were calculated from the reported bias measurements using the \cite{Tinker:2010} halo bias relation, and we assume bolometric corrections for the various wavebands given: 27 for 0.5-2 keV, and 20 for 2-10 keV \citep{Lusso:2012}\footnote{While the soft and hard band bolometric corrections are functions of luminosity, we use the empirically found values for $\sim 10^{12}~L_{\odot}$ from \cite{Lusso:2012} for simplicity. Using single bolometric corrections is fairly insensitive to lower luminosities, and results in conservative bolometric luminosity estimates for the luminous ($>10^{45.5}$ \ergs) sources.}. 
In the previous measurements, there seems to be a slight luminosity dependence, where moderate luminosity AGN typically reside in larger halos than luminous quasars. However, 
we find that the halo mass of our X-ray quasars are significantly higher than those found for optical quasars, by $\sim 0.5$ dex. This mass scale is more consistent with the halo masses of lower-luminosity X-ray-selected AGN; in particular, AGN at similar redshifts but with an order of magnitude lower luminosities from the COSMOS survey have similar estimated host halo masses as our high$-L$ bin \citep{Allevato:2011}. Meanwhile, optical quasars with comparable luminosities {\it and} redshifts (e.g., $L_{bol}\sim 10^{46}$ \ergs\ BOSS quasars at $z\sim 2.4$ \ergs\ and $L_{bol}\sim 3\times 10^{45}$ \ergs\ eBOSS quasars at $z\sim 1.7$; \citealt{Eftekharzade:2015,Laurent:2017}) have significantly lower host halo masses.  This suggests that biases in the different AGN/quasar selection methods, rather than luminosity differences, are the likely reason for the clustering differences between X-ray AGN and optical quasars. The potential causes for this bias are discussed in the following section.

Our lower-luminosity bin, on the other hand, resides in lower-mass halos consistent with the quasar and other X-ray AGN samples at similar luminosities and redshifts (e.g., quasars at $z\sim 1$, $\log~L_{bol}\sim 45$; \citealt{Laurent:2017}; X-ray AGN at $z\sim 0.8$, $\log~L_{bol}\sim 44.9$ and $z\sim 0.42$, $\log~L_{bol}\sim 45.2$; \citealt{Mountrichas:2016,Krumpe:2012}), although the errors on our measurement are large. This indicates that there are also redshift and luminosity dependencies on the effective host halo masses found for AGN, due to both the growth of structure over cosmic time, as well as the scaling relations between supermassive black holes and their galaxies. This is the likely reason that the typical host halo mass found for quasars in the COSMOS field \citep{Allevato:2016}, which have similar luminosities to our sample but are at higher redshifts ($z\sim 3.4$), are closer in mass to to the optical quasar hosts rather than other X-ray samples at $z=1-2$; there are far fewer halos of $10^{13}$ \Msun\ at $z=3.4$ than at $z=1.8$.

Recently, \cite{Jones:2019}  investigated the average AGN host halo masses as a function of bolometric (or X-ray) luminosity and redshift in a semi-numerical model of galaxy and black hole formation. They found that, for moderate luminosity-limited samples, there is a flat relation with halo mass due to the broad distribution of Eddington ratios. The relation steepens at high luminosities (i.e., halo mass increases with luminosity) since most of those objects are accreting at their Eddington limits. The halo masses also decrease with redshift at a given luminosity, since massive halos are rarer toward higher redshifts. These two competing effects dictate the relations between luminosity and redshift observed. Their findings at similar redshifts and luminosities are consistent with our results, to within error.

\subsection{Selection biases}

Luminous quasars from optical surveys have been found to reside in lower-mass halos than X-ray-selected AGN at $z=1-2$. If not because of luminosity dependencies, what are the other possible causes of this difference?

All AGN selection techniques are more likely to find AGN in higher-mass host galaxies for a given Eddington ratio distribution function, due to the scaling relations between black hole mass and stellar mass \citep{Aird:2012,Jones:2017,Azadi:2017}. If, for the same bolometric luminosity, X-ray AGN are \emph{more} biased toward being detected in large galaxies, then the clustering differences between X-ray AGN and optical quasars could be explained by the relationship between stellar mass and halo mass \citep[e.g.,][]{Behroozi:2013B}. This is feasible because optically-selected quasars typically require a higher contrast between the AGN point source and their host galaxy for detection, since the galaxy can more easily contaminate the AGN signal in the optical waveband. On the other hand, there is much less host galaxy contamination for AGN selected by X-rays, and so X-ray selection should be less sensitive to host galaxy stellar mass. Whether this is the case for this high-luminosity sample remains unclear, however, as stellar mass is difficult to estimate for luminous quasars. Recent work looking at AGN host galaxies selected by different techniques in the MOSDEF survey showed that optical and X-ray selections are similarly biased toward high stellar masses \citep{Azadi:2017}, although that study was based on two orders of magnitude fewer AGN than in the present work and used lower-luminosity sources. Using stellar mass estimates of the spectroscopic AGN sample in Stripe82X, we found that the clustering of our AGN were consistent with the prediction based on their stellar masses alone (see Figure \ref{fig:wp}). Comparing the X-ray-detected AGN in S82X with the fraction that were detected in SDSS ($\sim 20\%$), the median stellar mass of the X-ray AGN were indeed higher, but only by $\sim 0.1$ dex. It should be noted, however, that disentangling the host galaxy from the luminous quasar component is difficult, making the stellar mass estimates for those objects extremely uncertain.

Galaxy clustering also depends strongly on star formation rate \citep{Coil:2017} for a given stellar mass. Due to emission-line selection, optically-selected AGN are biased toward relatively lower star formation rates 
\citep{Trump:2015,Azadi:2017} with older stellar populations. However, since blue star-forming galaxies are less clustered than older, red galaxies, this would bias the clustering differences between optical and X-ray AGN in the opposite way as observed. Therefore, this bias is not the cause of the clustering differences between X-ray and optical AGN.

Lastly, it has been observed that AGN clustering depends on the obscuration of the nucleus, estimated either by absorbing column density \citep{Krumpe:2017,Powell:2018} measured from X-rays, or by IR color \citep{DiPompeo:2014,DiPompeo:2017}. Obscured AGN are typically found to be slightly more clustered than unobscured AGN, the reason for which is still not clear. Optical detection is less effective at finding absorbed AGN than X-rays, and so this could contribute to the observed difference. However, only $\sim 6\%$ of the high-luminosity XXM-XXL-north subsample have column densities over $10^{22}$ atoms/cm$^2$ (measured by their X-ray spectra). The majority of the S82X AGN also show broad lines in their optical spectra ($\sim 99\%$ for the high-luminosity bin) indicative of little nuclear obscuration, and therefore this may not be a large effect for this AGN sample, though we note that optical spectroscopy is biased toward bright (i.e., unobscured) AGN.

To summarize, the flux-limited samples typically used in X-ray AGN clustering analyses have different incompleteness compared to optical quasar samples. This incompleteness may vary over the redshift range, and affect the clustering amplitude found for a sample of a given luminosity. 
Selection effects driving the observed clustering differences was also concluded in \cite{Georgakakis:2018}, which reproduced the correlation functions of optical and X-ray AGN using semi-empirical simulations. This was done by assuming a single HOD and replicating the selections of each sample. Observationally, larger multiwavelength surveys are needed to fully characterize these selection effects. Only with larger, homogeneous samples can luminosity, redshift, and obscuration be independently controlled.

\subsection{Limitations of interpreting halo masses from large-scale bias}

The methods used for inferring a typical halo mass from the large-scale clustering strength of a sample of AGN rely on several assumptions. The first assumption is that the distribution of host halo masses is narrow. This could be valid if major mergers predominantly trigger AGN, as mergers prefer environments where the number density of galaxies is high, but where the relative velocities between them are sufficiently low \citep{Hopkins:2007}. However, many recent investigations have argued against major mergers being the main AGN triggering mechanism up to moderate redshifts for moderate-luminosity sources \citep[e.g.,][]{Simmons:2012,Kocevski:2012,Rosario:2015,Powell:2017,Hewlett:2017}, and even for the most luminous quasars \citep{Villforth:2014,Villforth:2017}. Additionally, investigations that have interpreted AGN clustering by forward modeling the AGN samples, by making simple assumptions and populating halo catalogs from $N-$body simulations, have argued that AGN reside in a wide range of environments with a broad distribution of host halo masses \citep{Powell:2018,Georgakakis:2018,Jones:2019}. Studies of black hole halo occupation in hydrodynamic simulations agree \citep{DeGraf:2017}. If this is the case, then the typical halo mass obtained from the bias may not represent the median halo mass hosting the AGN population due to incompleteness of the sample \citep{DeGraf:2017,Powell:2018}. Therefore, we caution against inferences made from host halo mass estimates derived from clustering bias.

An additional assumption is that halo clustering only depends on halo mass. From simulations it has been shown that there is an effect known as assembly bias \citep{Dalal:2008}, in which halos of the same mass cluster differently based on their formation epochs. Halos that have assembled their mass earlier in cosmic time cluster more strongly than halos formed later (which is related to the dependence of star formation rate/color on galaxy clustering; e.g., \citealt{Hearin:2013}). If any AGN property depends upon the halo assembly history, then the estimated halo mass from the AGN bias could be systematically incorrect. The clustering differences between obscured and unobscured AGN that have the same stellar mass distributions (and therefore presumably similar host halo mass distributions) have been suggested to be explained by this effect \citep{Powell:2018}. However, more investigation is needed to constrain the magnitude of assembly bias on observational AGN clustering measurements.

\subsection{Implications for the triggering mechanisms of X-ray luminous quasars}

Major mergers have been proposed to be a significant player in galaxy-AGN coevolution, especially for the most luminous quasars at moderate to high redshifts \citep[e.g.,][]{Hopkins:2006}. Theoretical models assuming major mergers are the dominant mechanism for igniting black hole accretion predict quasars to reside in halos of $\sim 4\times 10^{12}$ \Msun, corresponding to small group environments \citep{Hopkins:2008}. While this is typically found for luminous quasars ($L_{bol}=10^{45}-10^{47}$ \ergs) detected in optical bands from their large-scale clustering amplitude \citep[e.g.,][]{Ross:2009,Eftekharzade:2015,Laurent:2017}, we found higher halo masses for X-ray selected AGN with luminosities $L_{bol}\sim 10^{46}$ \ergs. If the host halo mass distribution of our sample is indeed narrow, then this typical halo mass value is inconsistent with triggering by predominantly major mergers. If instead the halo mass distribution is broad, such that this estimate is not representative of the typical halo mass in our sample, this also weakens the argument for major merger triggering since major mergers are most efficient in a narrow range of halo masses \citep{Hopkins:2007}. It is thus likely that secular, internal processes are still important even in high-luminosity AGN, although more studies of the clustering properties of merging galaxies are needed. 
\vspace{8mm}

\section{Summary}
In this study we measured the clustering of X-ray-selected quasars in the Stripe 82X and XMM-XXL-north surveys, which span a combined area of 38 deg$^{2}$ (including large contiguous areas only). We specifically looked for any luminosity dependence in the inferred host halo masses of accreting supermassive black holes.

We found that the AGN in our higher-luminosity/redshift bin ($\log L_{X}\geq 44.5$ [\ergs]) reside in larger-mass halos ($\log M_{h}=13.0 \pm 0.2$ [\Msun]) than for our lower-luminosity/redshift AGN ($\log M_{h}=12.4^{+0.4}_{-0.8}$ [\Msun]; $43\leq \log L_{X}<44.5$ [\ergs]), inferred from their large-scale clustering bias. While not very significant, this goes in the opposite direction than found in several previous studies.

The typical host halo mass of $\sim 10^{13}$ \Msun\ measured for the $L_{X}\sim 10^{45}$ \ergs AGN at $z\sim 1.8$ is consistent with previously estimated halo masses hosting less luminous X-ray-selected AGN at similar redshifts, while being larger than those hosting optically-selected quasars of similar luminosities and/or redshifts. We argue that selection biases drive the differences in the clustering bias found for various AGN samples, as well as more complicated dependencies on luminosity and redshift; differences of this magnitude are easy to bridge depending on systematics that are not presently controlled for. Larger homogeneous samples across wide ranges of redshift and luminosity are needed to disentangle these effects, which will be possible with future surveys like eROSITA. Future work characterizing the host galaxy and AGN properties of this high-luminosity AGN sample will also elucidate these biases, and will help determine the dominant parameters on which AGN clustering depends.

\acknowledgments
We thank the referee for helpful comments that improved the paper. M.P. also thanks Antonis Georgakakis, Viola Allevato, Mackenzie Jones, Sarah Eftekharzadeh, Manodeep Sinha, and Steven Murray for useful discussions. The authors would like to acknowledge support for this work through NSF grant 1715512, NASA ADAP grant 80NSSC18K0418, NASA-SWIFT GI: Nr. 80NSSC18K0505, and Yale University.

\software{CorrFunc \citep{Sinha:2017},
HMF \citep{hmf:2014}, halomod \citep{halomod}, 
Astropy \citep{Astropy:2013}, Halotools \citep{Hearin:2017},
Matplotlib \citep{matplotlib:2007}, Cigale \citep{Burgarella:2005,Serra:2011}.}

\bibliographystyle{yahapj}
\bibliography{references}

\appendix
We used the {\tt Cigale} software (Code Investigating GALaxy Emission; \citealt{Burgarella:2005,Noll:2009,Serra:2011,Ciesla:2015,Boquien:2019}) to fit the spectral energy distributions of the AGN in S82X with spectroscopic redshifts. We input the fluxes of the multiwavelength data (see \citealt{Ananna:2017}) and their redshifts to obtain the estimated host galaxy stellar masses and their uncertainties. The parameters and their ranges assumed for the fitting procedure are given in Table \ref{tb:t3}.  

\begin{table*}[h]
\caption{Models and parameter ranges used in the {\tt Cigale} SED fitting.} 
\centering
\setlength{\tabcolsep}{0.1mm}
\begin{tabular}{cc}
       \hline
{\bf Parameter} &  {\bf Model/values} \\
        \hline
        \hline
\multicolumn{2}{c}{{\bf \cite{Maraston:2005} stellar population synthesis model}} \\
initial mass function & Salpeter \\
metallicity & 0.02 \\
\hline
\multicolumn{2}{c}{{\bf Delayed Star Formation History model}} \\
$\tau$ of stellar population models (Myr) & 500, 1000, 3000, 5000, 10000 \\ 
Age (Myr) & 4000, 5000, 5500 \\
\hline
\multicolumn{2}{c}{{\bf \cite{Calzetti:2000} dust extinction}} \\
reddening E(B-V) young & 0.05, 0.1, 0.15, 0.2, 0.25, 0.3, 0.35, 0.4, 0.5, 0.6 \\ 
E(B-V) reduction factor between old and young stellar population & 0.44 \\
\hline
\multicolumn{2}{c}{{\bf \cite{Dale:2014} dust template}} \\
IR powerlaw slope &  1.5,2.0,2.5 \\
\hline
\multicolumn{2}{c}{{\bf \cite{Fritz:2006} model for AGN emission}} \\ 
ratio between outer and inner dust torus radii &   30, 100 \\
9.7 $\mu m$ equatorial optical depth & 0.3, 3.0, 6.0, 10.0 \\
Parameter for radial dust distribution in torus ($\beta$) & -0.5 \\
Parameter for angular dust distribution in torus ($\gamma$) & 0.0, 2.0, 6.0 \\
Opening angle of the torus ($\Theta$) & 100 \\
Line of sight angle ($\Psi$) & 0.001, 50.100, 89.990 \\
$L_{IR}$ AGN fraction& 0.0, 0.05, 0.1, 0.15, 0.2, 0.25, 0.3, 0.4, 0.5, 0.6, 0.7, 0.8 \\
\hline
\label{tb:t3}
\end{tabular}
\end{table*}

\end{document}